\pdfoutput=1
\documentclass[aps,prd,tightenlines,floatfix,twocolumn,superscriptaddress,showpacs]{revtex4}
\usepackage{graphicx}
\usepackage{wrapfig}
\usepackage{hyperref}
\usepackage[all]{xy}
\usepackage[usenames]{color}

\newcommand{\nc}{\newcommand}
\nc{\beq}{\begin{equation}}
\nc{\eeq}{\end{equation}}
\nc{\bea}{\begin{eqnarray}}
\nc{\eea}{\end{eqnarray}}
\nc{\n}{\nonumber \\}

\nc{\physrep}{Physics Reports}

\nc{\sig}{\langle \sigma_{\rm a} v \rangle}

\begin{document}  

\title{Probing the Small Scale Matter Power Spectrum 
through Dark Matter Annihilation in the Early Universe}

\author{Aravind Natarajan}
\email{anat01@me.com}
\affiliation{Kavli Institute for the Physics and Mathematics of the Universe (WPI), The University of Tokyo Institutes for Advanced Study, 
Kashiwa-no-ha, Chiba, 277-8583, Japan} 
\affiliation{Department of Physics, Engineering Physics, and Astronomy, Queen's University, Kingston, Ontario K7L 3N6, Canada}
\affiliation{Department of Physics and Astronomy, University of Pennsylvania, 209 South 33rd Street, Philadelphia, PA 19104, USA}
 
\author{Nick Zhu}
\affiliation{Department of Physics, University of Tokyo, Bunkyo, Tokyo 113-0033, Japan} 
\affiliation{Department of Astrophysical Sciences, Princeton University, Princeton, NJ 08544}

\author{Naoki Yoshida}
\affiliation{Kavli Institute for the Physics and Mathematics of the Universe (WPI), The University of Tokyo Institutes for Advanced Study, 
Kashiwa-no-ha, Chiba, 277-8583, Japan} 
\affiliation{Department of Physics, University of Tokyo, Bunkyo, Tokyo 113-0033, Japan}

\begin{abstract}
\noindent Recent observations of the cosmic microwave background (CMB)
anisotropies and the distribution of galaxies, galaxy clusters, 
and the Lyman $\alpha$ forest have constrained the shape of the power 
spectrum of matter fluctuations on large scales $k \lesssim$ few $h$/Mpc. 
We explore a new technique to constrain the matter power 
spectrum on smaller scales, assuming the dark matter is a Weakly Interacting 
Massive Particle (WIMP) that annihilates at early epochs. 
Energy released by dark matter annihilation can modify the spectrum 
of CMB temperature fluctuations and thus CMB experiments such as Planck  
have been able to constrain the quantity $f \sig /m_\chi \lesssim 1/88$ 
picobarn$\times c/$GeV, where $f$ is the fraction of energy absorbed 
by gas, $\sig$ is the annihilation rate assumed constant, 
and $m_\chi$ is the particle mass. We assume the standard scale-invariant 
primordial matter power spectrum of $P_{\rm prim}(k) \sim k^{n_s}$ 
at large scales $k < k_{\rm p}$, while we adopt the 
modified power law of $P_{\rm prim}(k) \sim k_p^{n_{\rm s}} (k/k_p)^{m_{\rm s}}$
at small scales. We then aim at deriving constraints on
$m_{\rm s}$. For $m_{\rm s} > n_{\rm s}$,
the excess small-scale power results 
in a much larger number of nonlinear small mass halos, 
particularly at high redshifts.  
Dark matter annihilation in these halos releases sufficient energy 
to partially ionize the gas, and consequently modify the spectrum of CMB 
fluctuations. 
We show that the recent Planck data can already be used to constrain 
the power spectrum on small scales. 
For a simple model with an NFW profile with halo 
concentration parameter $c_{200}$ = 5 
and $f \sig /m_\chi$ = 1/100 picobarn$\times c/$GeV, we can limit 
the mass variance $\sigma_{\rm max} \lesssim 100$ at the 95\% confidence level, 
corresponding to a power law index $m_{\rm s} < 1.43 (1.63)$ for $k_{\rm p}$ 
= 100 (1000) $h$/Mpc. Our results are also relevant to theories that 
feature a running spectral index.
\end{abstract}

\maketitle

\section{Introduction}
The nature of dark matter is unknown and remains 
 one of the greatest mysteries in astrophysics and cosmology. 
Weakly Interacting Massive Particles (WIMPs) are one of the leading candidates 
for the dark matter of the Universe, and large experiments are 
being conducted to detect WIMP dark matter through direct, indirect, 
and collider experiments. Some direct detection experiments such 
as DAMA \cite{dama} have observed an annual modulation consistent 
with the presence of dark matter particles of mass $8-15$ GeV 
interacting with a spin-independent cross section of $0.01-0.1$ femtobarn. 
Recent observations of the Milky Way center \cite{gal_center} by the Fermi 
gamma ray telescope also seem to indicate an excess of gamma rays, 
consistent with dark matter particles of mass $m_\chi = 31-40$ GeV annihilating 
at a rate $\sig = (1.4 - 2.0) \times 10^{-26}$ cm$^3/$s. 
These exciting results are, however,  inconsistent with the XENON \cite{xenon} 
and LUX experiments \cite{lux}, and are also disfavored by the non-detection 
of dark matter annihilation in the local dwarf galaxies \cite{dwarf1,dwarf2}. 

The cosmic microwave background has been shown to be an excellent probe of 
WIMP dark matter annihilation at high redshifts \cite{cmb1,cmb2,arvi1,arvi2,arvi3}. 
Particle annihilation releases and injects energy into the cosmic diffuse gas, 
resulting in both ionization and heating. Free electrons scatter CMB photons 
and cause damping in the temperature anisotropy power spectrum 
at intermediate and small angular scales.
Also the CMB polarization power spectrum is boosted at very large scales. 
Precise measurement of the CMB by Planck, WMAP, ACT, and SPT have 
already placed
tight constraints on dark matter properties\cite{cmb3,cmb4,cmb5,cmb6,cmb_arvi}.  
Interestingly, the recent results from the Planck collaboration \cite{newplanck_cosm} 
constrain the annihilation parameter $p_{\rm ann} = f\sig/m_\chi < 3.4 \times 10^{-28}$ cm$^3$s$^{-1}$GeV$^{-1}$ = $(1/88.3)$ pb$\times$c$/$GeV, 
where $f$ is the fraction of energy absorbed by gas. 
For realistic values of $f \approx 0.35$ for the $b \bar b$ channel \cite{slatyer}, 
and $\sig = 0.727$ pb$\times$c \cite{arvi_mssm}, 
one obtains a bound on the dark matter mass $m_\chi > 22.4$ GeV 
at the 95\% confidence level\footnote{It must be noted that the CMB bound only 
applies for $s$-wave (velocity independent) annihilation.}.

The CMB constraints on dark matter particle properties are
derived on the assumption that only
annihilation of free `unbound' particles contribute to the net energy release, 
i.e. annihilation of particles  
bound in nonlinear objects, ``dark halos'', is not considered. 
The assumption is appropriate
because, in the standard cosmology, 
nonlinear objects appear only relatively late, at redshifts 
$z \sim 30-50$. Then the gas density is already small, 
and CMB photons do not interact with free electrons 
unless the gas is significantly ionized. 
Nonlinear halos would be, however, important if they formed much earlier, 
i.e. at redshifts $z > 100$. Such a case is possible if the primordial
density fluctuations have some excess power at small length scales.

\begin{figure*}[!t]
\begin{center}
\scalebox{0.9}{\includegraphics{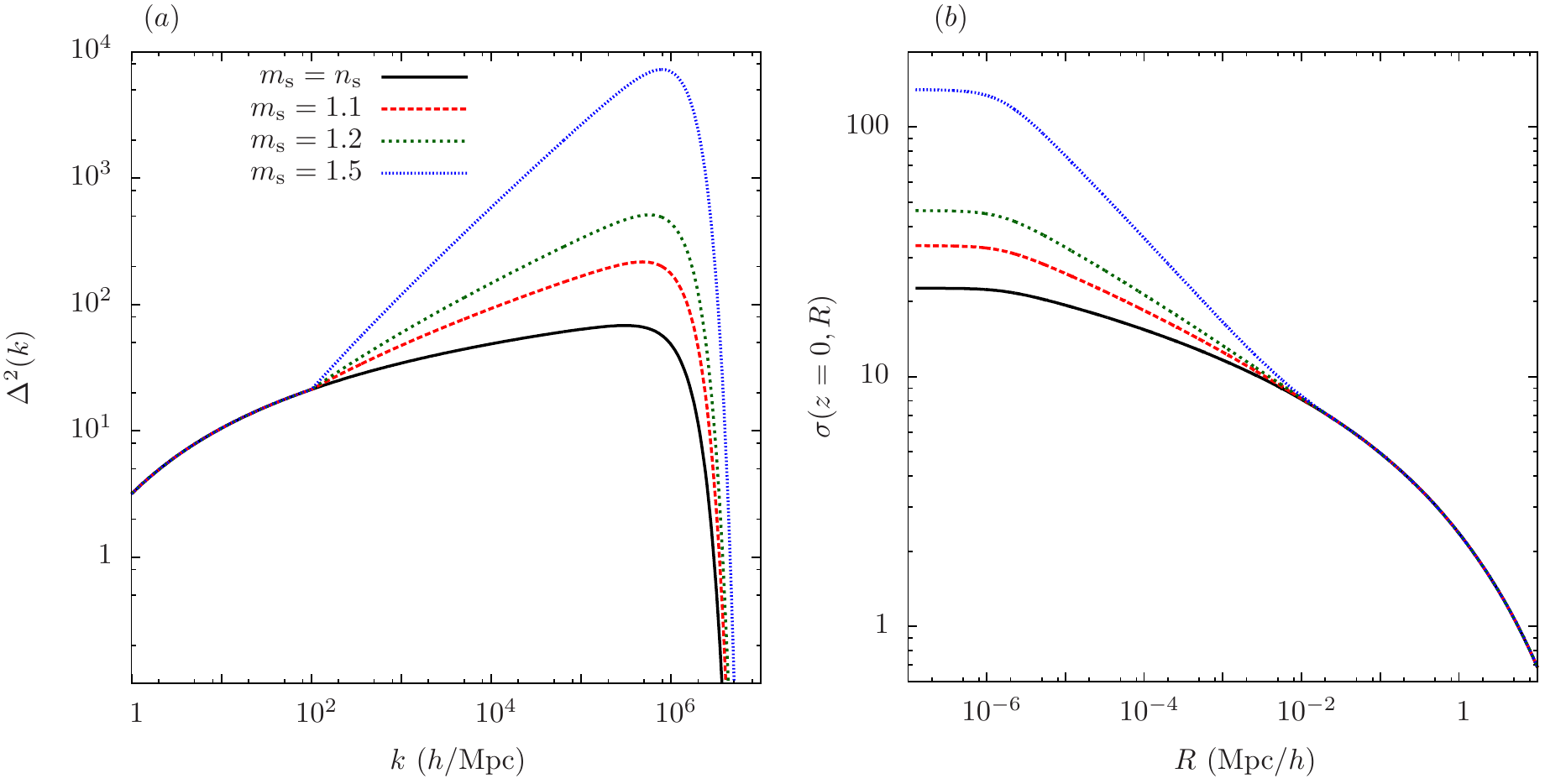}}
\end{center}
\caption{ We plot the dimensionless matter power spectrum 
for the standard cosmology (solid, black), as well as for the 
cases $m_{\rm s}$ = 1.1, 1.2, and 1.5, for $k_{\rm p}$ = 100 $h$/Mpc
(Panel a). 
An exponential cut-off is applied at the free streaming 
scale $k_{\rm fs}$ chosen to be  $ 10^6$  $h$/Mpc. Panel (b) 
shows the corresponding 
standard deviation $\sigma$ of fluctuations normalized to $\sigma_8 = 0.8$.
\label{fig1} }
\end{figure*}

The simplest inflationary theories predict a nearly scale invariant 
primordial curvature power spectrum $P_{\mathcal{R}} \sim k^{n_{\rm s}-1}$, 
resulting in a present day matter power spectrum:
\bea
P_{\rm m}(z,k) &\propto& A \, k^{n_{\rm s}} T^2(k) \frac{D^2(z)}{D^2(0)} \n
&=& P_{\rm prim}(k) T^2(k) \frac{D^2(z)}{D^2(0)},
\eea
where $P_{\rm prim}(k) \propto k^{n_{\rm s}}$ is the primordial matter power spectrum 
on very large scales $k < 10^{-3}$ $h$/Mpc, 
$T(k)$ is the transfer function, and $D(z)$ is the growth factor.  
Data from the Wilkinson Microwave Anisotropy Probe (WMAP) and the Atacama 
Cosmology Telescope (ACT) was used by \cite{hlozek} to reconstruct the 
matter power spectrum at wavenumbers $0.001 < k < 0.19$ Mpc$^{-1}$. 
On smaller length scales, one may use the Sloan Digital Sky Survey measurements 
of the clustering of galaxies to probe scales up to $k \sim 0.2$ $h$/Mpc.
On even smaller scales, the flux power spectrum of 
the Lyman $\alpha$ forest may be used to probe 
the matter power spectrum for $k < 2$ Mpc$^{-1}$.
Unfortunately, at $k \gtrsim 10$ $h$/Mpc, there exist no direct observations 
of the matter power spectrum. Possible probes of the small scale  
power spectrum include the use of Type Ia supernova lensing dispersion \cite{sn1a},  
ultra compact mini halos \cite{ucmh1,ucmh2,ucmh3,ucmh4}, 
and the dissipation of acoustic waves by Silk damping \cite{silk1,silk2}. 

In the present paper, we explore a new probe of the primordial
density fluctuations on very small scales. 
Let us consider a simple power-law for the matter power spectrum 
at $k > k_{\rm p}$:
\bea
P_{\rm prim} &=& A \, k^{n_{\rm s}} \;\;\;\;\;\;\;\;\;\;\;\;\;\;\;\;\;\;\;\; k \leq k_{\rm p} \n
&=& A \, k_{\rm p}^{n_{\rm s}} \, \left ( k/k_{\rm p} \right )^{m_{\rm s}}  \;\;\;\;k > k_{\rm p},
\label{mod}
\eea
which is consistent with all available observations provided 
the pivot wavenumber $k_{\rm p}$ 
is large enough, 
say $k_{\rm p} \gtrsim 10$ $h$/Mpc. Essentially, we examine if there is excess
power on the relevant small length scales through energy injection from
dark matter annihilation in the early universe. The basic idea is as follows.
The mass variance  $\sigma^2(z,M)$ is computed by integrating the dimensionless 
power spectrum $\Delta^2(k) = k^3 P(k) / 2 \pi^2$ over a window function: 
\beq
\sigma^2 (z,M)= \frac{D^2(z)}{D^2(0)} 
\int \frac{{\rm d}k}{k} \, \frac{k^3 P(k)}{2 \pi^2} W^2(kR).
\label{sig}
\eeq
The normalization constant $A$ is chosen such that $\sigma_8 = 0.8$, 
where $\sigma_8$ is the root mean square mass fluctuation in a sphere 
of radius 8 Mpc/$h$. The linear growth function of matter overdensities
is denoted by $D(z)$.
With the Planck cosmology, 
we find $D(z=0)$ = 0.757. 
The window function $W(kR)$ for comoving scale $R$ is 
conveniently given by
\beq
W(x) = \frac{3 \left[ \sin x - x\cos x \right ] }{ x^3},
\eeq
and $R$ is the comoving radius that encloses a mass $M$. 
Fig. \ref{fig1}(a) shows the matter power spectrum for the standard 
cosmology (solid lines, black) 
computed using the Eisenstein-Hu transfer 
function \cite{eisenstein_hu1, eisenstein_hu2}. 
Also plotted are curves 
for $n_{\rm s}$ = 0.96, and the modified power law of Eq.~\ref{mod}, 
for $m_{\rm s}$ = 1.1, 1.2, and 1.5. An exponential cutoff is imposed 
at the free streaming scale $k_{\rm fs} \sim 10^6$ $h$/Mpc \cite{fs1,fs2,fs3}, 
to account for the finite velocity dispersion of WIMP dark matter, 
and to make the integral in Eq.~\ref{sig} finite. This ensures that there 
is a minimum halo mass $M_{\rm min} \sim 10^{-6} M_\odot$. 
Panel (b) shows the 
standard deviation $\sigma$ of density fluctuations, for these
models. Note that $\sigma_{\rm max} = \sigma(M_{\rm min})$ is very 
sensitive to $m_{\rm s}$, although $\sigma_8$ = 0.8 for all models.

\section{Dark matter annihilation in halos}

Consider an overdensity of weakly interacting dark matter particles.  
The number of WIMPs in a volume $\delta V$ is $(\rho_\chi/m_\chi) \, \delta V$, 
where $\rho_\chi$ is the density of WIMPs, and the probability of WIMP 
annihilation in a time $\delta t$ is $\sig \delta t$. 
The number of WIMP annihilations per unit time per unit volume is 
then equal to $\sig \rho^2_\chi / m^2_\chi$. Since each annihilation 
releases $m_\chi$ of energy per particle, 
the total energy released per unit time 
per unit volume equals 
\beq
\frac{{\rm d}E}{{\rm d}t {\rm d}V} = \frac{\sig}{m_\chi} \, \rho^2_\chi.
\label{dedtdv_free}
\eeq

The energy  per unit time due to particle annihilation 
in a bound halo of radius $r_{200}$ is obtained 
by integrating Eq.~\ref{dedtdv_free} 
over the halo volume:
\beq
\frac{dE}{dt} = \frac{\sig}{m_\chi} \, \int_0^{r_{200}} dr \, 4 \pi r^2 \rho_{\rm halo}^2(r).
\label{dedt}
\eeq
$r_{200}$ is the radius at which the mean density enclosed equals 200 
times the cosmological mean at the redshift of formation of the halo:
\beq
\frac{3M}{4\pi r^3_{200}} = 200 \rho_0 [1+z_{\rm f}(M)]^3,
\eeq
where $\rho_0$ is the mean dark matter density at the present epoch ($z=0$), 
and $z_{\rm f} (M)$ is the formation redshift 
of a halo of mass $M$. 
We may parameterize the halo density profile as:
\beq
\rho(x) = \frac{\rho_{\rm s}}{x^\alpha (1+x)^\beta},
\label{nfw}
\eeq
where $x = r / r_{\rm s}$ is a dimensionless radius. $\rho_{\rm s}$ and $r_{\rm s}$ are constants for a halo. The well known Navarro-Frenk-White (NFW) \cite{nfw} form is 
obtained when we set $\alpha = 1$ and $\beta = 2$. 

\begin{figure*}[!t]
\begin{center}
\scalebox{0.85}{\includegraphics{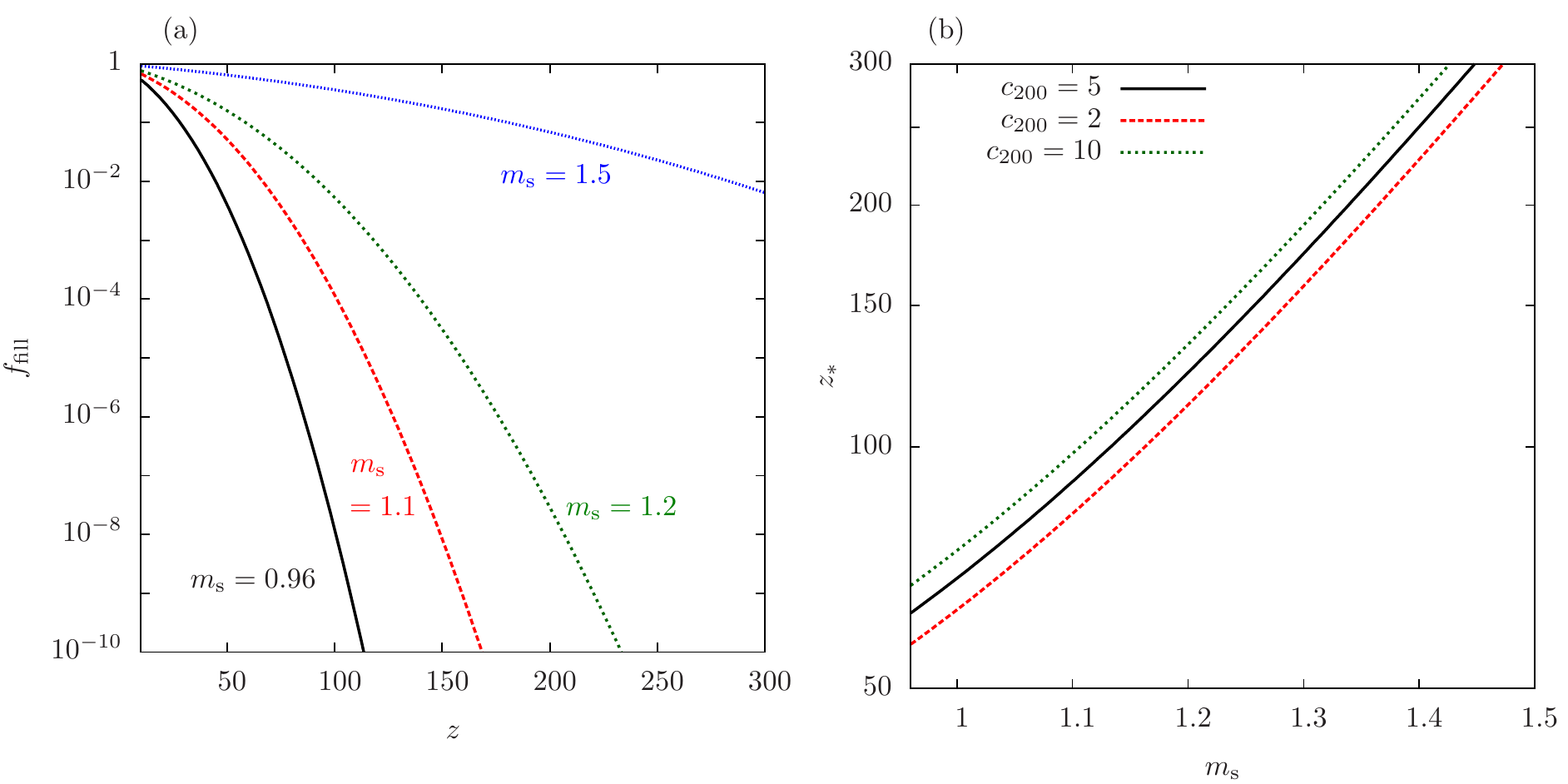}}
\end{center}
\caption{We plot 
the filling fraction $f_{\rm fill}$ for different values of $m_{\rm s}$, 
for $k_{\rm p} = 100$ $h$/Mpc (Panel a). 
The difference appears small at $z=0$, but is substantial for 
large $z$. Panel (b) shows the redshift $z_\ast$ below which 
the halo contribution exceeds the free particle contribution. 
It is not very sensitive to the concentration parameter 
because of the exponential decrease of $f_{\rm fill}$ with $z$.
\label{fig2} }
\end{figure*}

We define the concentration parameter as:
\beq 
c_{200} = \frac{r_{\rm 200} }{ r_{\rm s} }.
\label{conc}
\eeq
Note that we have defined $r_{200}$ and $c_{200}$ at the 
formation epoch. 
We may now express $\rho_{\rm s}$ and $r_{\rm s}$ in terms 
of $M$ and $c_{200}$. Then the rate of energy release (Eq.~\ref{dedt}) is
\beq
\frac{{\rm d}E_{\rm halo}}{{\rm d}t} = \frac{\sig}{m_\chi} \, \frac{200}{3} M \rho_0 [1+z_{\rm f}(M)]^3 \,   f_{\rm conc}(c_{200}),
\label{halo}
\eeq
where $f_{\rm conc}(c_{200})$ is calculated for the density profile 
with concentration parameter $c_{200}$ as
\beq
f_{\rm conc} = \frac{c^3_{200} \int_\epsilon^{c_{200}} 
{\rm d}x \, x^{2-2\alpha} (1+x)^{-2\beta}}
{\left[ \int_0^{c_{200}} {\rm d}x \, x^{2-\alpha} (1+x)^{-\beta}\right ]^2 }.
\label{gamma}
\eeq
The lower limit $\epsilon$ is required when the index $\alpha > 1.5$. 
For the NFW profile with $\alpha=1$ and $\beta=2$, we can set $\epsilon = 0$, 
and then Eq.~\ref{gamma} is integrated analytically to yield
\beq
f_{\rm conc} = \frac{c^3_{200}}{3}\frac{1 - (1+c_{200})^{-3}}{ \left[\ln(1+c_{200}) - c_{200}(1+c_{200})^{-1} \right ]^2}.
\eeq
A halo of mass $\sim 10^{12} M_\odot$ similar to the Milky Way 
is expected to have a concentration parameter $c_{200} \sim 10$ \cite{maccio}. 
Earth-mass microhalos, on the other hand, are not expected to have 
large concentration parameters. 
\cite{diemand} found concentration parameters $c_{200} \lesssim 3$ 
for such very small halos.  
We simply assume a constant $c_{200}$ = 5 independent of mass.

To evaluate Eq.~\ref{halo}, we need to determine the redshift of formation 
of the halo. Following the Press-Schechter formalism \cite{ps}, 
we assume that the probability of finding a halo of mass $M$ at 
a redshift $z$ is $\propto \exp -[\delta_{\rm c}^{2}/2\sigma^{2} (M,z)]$, 
where $\delta_{\rm c} = 1.686$ is the threshold for halo formation 
in linear theory. We can then estimate the averaged quantity:
\beq
\langle \left [ \frac{1+z_{\rm f}(M)}{1+z} \right ]^3  \rangle 
=  \frac{ \frac{1}{x^3_\ast}\int_{x_\ast}^\infty {\rm d}x \, 
x^3 e^{-x^2}}{\int_{x_\ast}^\infty  {\rm d}x \,  e^{-x^2}},
\label{av}
\eeq
where $x_\ast = \delta_{\rm c}/\sqrt{2}\sigma(z,M)$. 
Note that Eq.~\ref{av} approaches unity for large halo masses and large redshifts.

The total energy due to WIMP annihilation per unit time and 
per unit volume may be obtained by integrating Eq.~\ref{halo} 
over the halo distribution:
\bea
\frac{{\rm d}E}{{\rm d}t {\rm d}V} 
&=& \left(1 + z \right )^3 \int_{M_{\rm min}}^\infty {\rm d}M \, 
\frac{{\rm d}N}{{\rm d}M} \, \frac{{\rm d}E_{\rm halo}}{{\rm d}t} \n
&=& \frac{\sig}{m_\chi} \, \frac{200  \rho_0}{3} f_{\rm conc}(c_{200}) (1+z)^6 \n
&\times& \int_{M_{\rm min}}^\infty {\rm d}M \, M \, 
\frac{{\rm d}N}{{\rm d}M} \, \langle \left [ \frac{1+z_{\rm f}(M)}{1+z} \right ]^3  \rangle
\label{dedtdv}
\eea

The comoving number density of halos
is calculated from, for instance,
the Press-Schecter mass function as
\beq
\frac{{\rm d}N}{{\rm d}M} = \sqrt{\frac{2}{\pi}} \rho_0 \left| \frac{1}{\sigma} \frac{{\rm d}\sigma}{{\rm d}M} \right | \frac{\delta_{\rm c}}{\sigma} \exp \left [ -\frac{\delta_{\rm c}}{2\sigma}\right ]^2.
\eeq
We have simplified the mass function by 
setting $\rho_0$ equal to the dark matter density, 
rather than the matter density. 
Let us define the filling factor $f_{\rm fill}(z)$ 
as the fraction of matter in bound halos:
\bea
f_{\rm fill}(z) &=& \frac{1}{\rho_0} \, 
\int_{M_{\rm min}}^\infty {\rm d}M M \, \frac{{\rm d}N}{{\rm d}M} \n
&=& {\rm erfc} \left[ \frac{\delta_{\rm c} D(0) (1+z)}{\sqrt{2} \, \sigma(0,M)} \right ],
\eea
We also define the quantify $\zeta(z)$ by 
\beq
f_{\rm fill}(z) \zeta(z) = \frac{1}{\rho_0}  \int_{M_{\rm min}}^\infty dM M \, \frac{{\rm d}N}{{\rm d}M} \langle \left [ \frac{1+z_{\rm f}(M)}{1+z} \right ]^3  \rangle. 
\eeq
which is larger than 1 at low redshifts 
but approaches 1 for large $z$. 
Fig.~\ref{fig2}(a) shows the filling fraction for different choices of $m_{\rm s}$.
We set $k_{\rm p} = 100$ $h$/Mpc as our fiducial model parameter. 
The black curve shows the case $m_{\rm s} = n_{\rm s} = 0.96$, 
i.e. the standard power law.  
Clearly, the filling factor can be many orders of magnitude larger
at high redshifts if $m_{\rm s} > n_{\rm s}$.

It is also interesting to compute $z_\ast$, the redshift below 
which the nonlinear halo contribution exceeds the free particle 
contribution. 
Fig.~\ref{fig2}(b) shows $z_\ast$ calculated
for different values of the 
concentration parameter.¡¡It is not very sensitive to 
the concentration parameter due to the exponential decrease 
of the mass function.  For the standard matter power spectrum, 
halos are only important at redshifts $z \lesssim 50$. 
However, for $m_{\rm s} = 1.5$, halos contribute significantly 
even at $z=300$. 

At high redshifts, we may make the ``on the spot'' approximation
when calculating the net energy input to the gas. 
Namely, we may safely ignore the propagation of high energy annihilation
products from the redshift of emission 
to the redshift of absorption. 
The net energy absorbed per atom per unit time at a redshift $z$ is given by:
\bea
\xi(z) &=& \frac{f}{n_{\rm b}(z)} \frac{{\rm d}E}{{\rm d}t {\rm d}V} \n
&=& \frac{f\sig \bar m}{m_\chi} \left (\frac{\rho_{\rm crit}}{h^2} \right )\frac{\left( \Omega_\chi h^2 \right )^2}{\Omega_{\rm b}h^2} (1+z)^3 \n
&\times& \left[ 1-f_{\rm fill}(z) + \frac{200}{3} f_{\rm conc}(c_{200}) f_{\rm fill}(z) \zeta(z) \right ],
\eea
where $f$ is the fraction of energy absorbed by the gas, $n_{\rm b}(z)$ is the baryon number density, and $\bar m$ is the mean nucleon mass, assuming 76\% hydrogen 
and 24\% helium. $\Omega_{\rm b}$, $\Omega_\chi$, and $\Omega_{\rm m}$ are the baryon, dark matter, and total matter fractions at the present epoch.

\begin{figure*}[!t]
\begin{center}
\scalebox{0.75}{\includegraphics{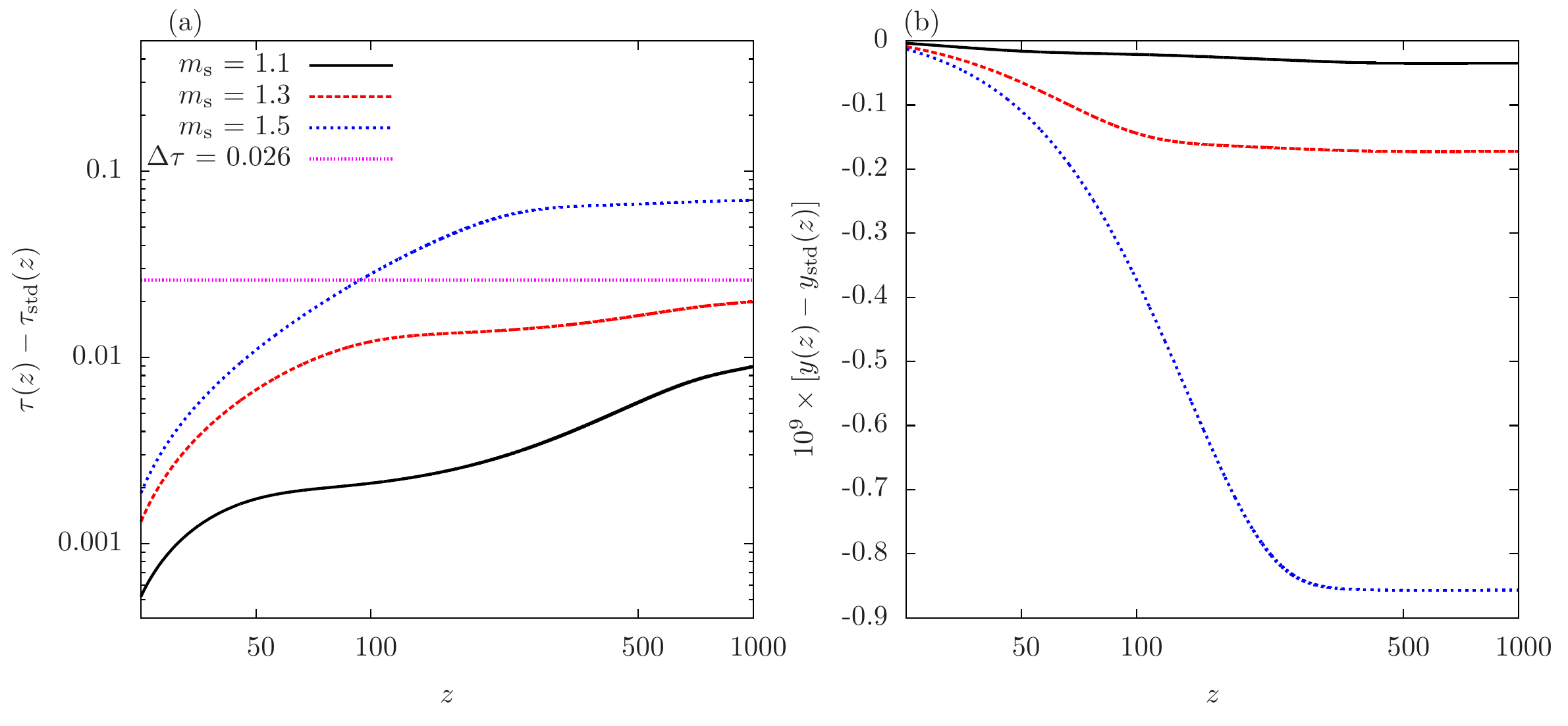}}
\end{center}
\caption{ The excess optical depth due to dark matter annihilation
including the contribution from nonlinear haloes, 
for $m_{\rm s} = 1.1, 1.3, 1.5$ (Panel a). 
The magenta line shows $\Delta\tau = 0.026$ which is the maximum 
allowed excess optical depth above $z=6$ (see text). 
Panel (b) shows the corresponding excess Compton $y$ parameter due to dark matter
annihilation as a function of $z$.
\label{fig3} }
\end{figure*}

\begin{figure*}[!t]
\begin{center}
\scalebox{0.75}{\includegraphics{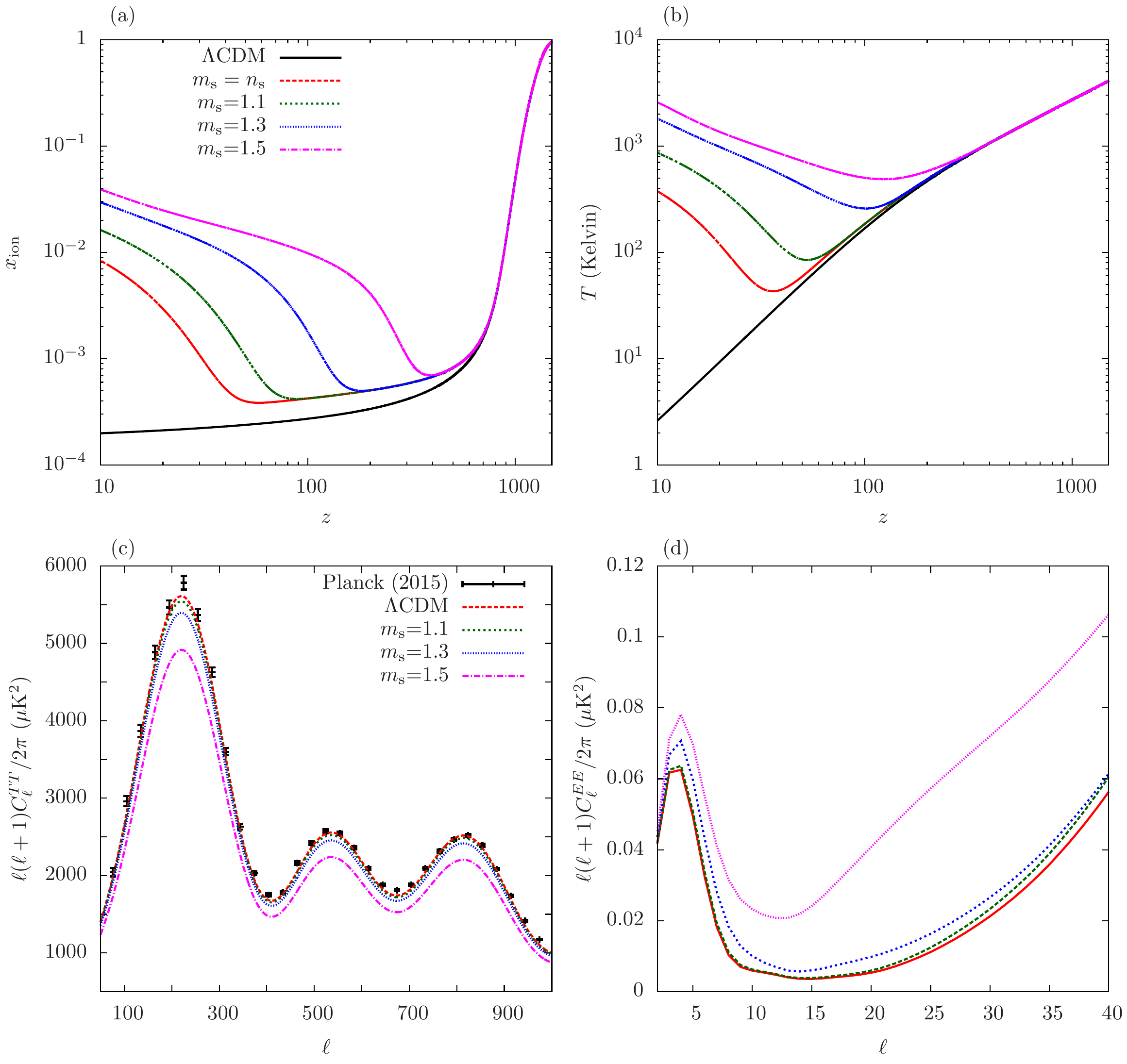}}
\end{center}
\caption{ The top panels (a) and (b) show the ionization 
and temperature history of the Universe. The black curve is 
plotted for the standard $\Lambda$CDM cosmology, i.e. ignoring 
dark matter annihilation. The red curve ($m_{\rm s} = n_{\rm s}$) 
is for a standard power spectrum, but includes the effect of 
dark matter annihilation. The green, blue, and magenta curves 
are plotted for the modified power law of Eq. \ref{mod}. 
The bottom panels (c) and (d) show the  temperature and 
polarization power spectra for the four models. 
\label{fig4} }
\end{figure*}

\section{CMB constraint}

In the previous section, we computed the rate of energy release 
due to dark matter annihilation in halos. 
The fraction $f$ of the released energy is absorbed 
by the gas, and a fraction $\eta_{\rm ion}(x_{\rm ion})$ of 
this energy goes into ionization, 
whereas a fraction $\eta_{\rm heat}(x_{\rm ion})$ goes into heating. 
We use the results obtained by \cite{furl_stoever} to estimate 
$\eta_{\rm ion}$ and $\eta_{\rm heat}$. 
The ionization and temperature evolution follow the equations \cite{cmb_arvi}:
\bea
-(1+z) H(z) \frac{{\rm d}x_{\rm ion}(z)}{{\rm d}z}  &=& \mu \left [1-x_{\rm ion}(z) \right ] \eta_{\rm ion} (z)  \xi(z) \n
  &-& n(z) x^2_{\rm ion}(z) \alpha(z)  \n
-(1+z) H(z) \frac{{\rm d}T(z)}{{\rm d}z}  &=& - 2T(z) H(z)  + \frac{2 \eta_{\rm heat}(z)} {3 k_{\rm b}} \, \xi(z)  \n
&+&   \frac{ x_{\rm ion}(z) \left [ T_\gamma(z) - T(z) \right ]}{t_{\rm c} (z)}.
\label{ion_T}
\eea
$\mu \approx 0.07$ eV$^{-1}$ is the inverse of the average ionization 
energy per atom, 
neglecting double ionization of helium. 
In the above equations, $\alpha$ is the case-B recombination coefficient, $T_\gamma$ is the CMB temperature, $k_{\rm b}$ is Boltzmann's constant, and $t_{\rm c}$ is the Compton cooling time scale $\approx$ 1.44 Myr $[30/(1+z)]^4$. 
The last term in the temperature evolution equation accounts for the transfer of energy between free electrons and the CMB by Compton 
scattering \cite{coupling, recfast1, recfast2}. 
In the temperature coupling term,
we assume $x_{\rm ion} \ll 1$ and ignore the helium number fraction. 
In practice, we compute $x_{\rm ion}$ and $T_{\rm gas}$ using a modified version of the publicly available {\scriptsize RECFAST} program \cite{recfast1, recfast2}.

CMB photons scatter off free electrons that are present due to partial ionization of the gas. Thomson scattering of the CMB causes 
damping of the temperature anisotropy $TT$ power spectrum, 
as well as a boost in the large angle $EE$ 
polarization power spectrum \cite{cmb_arvi}. 
The scattering is quantified by means of the optical 
depth defined as the scattering cross section times 
the free electron density integrated along the line of sight:
\bea
\tau(z_1,z) &=& \int {\rm d}t \, c \,  \sigma_{\rm T} n_{\rm e}(z) 
\label{opt_depth}
\eea
where $\sigma_{\rm T}$ is the Thomson cross section.
We also calculate the `excess' contribution as
\bea
\tau - \tau_{\rm std} &=& \frac{c \sigma_{\rm T} \left( \rho_{\rm crit}/h^2 \right )}{H_{100} \, \bar m} \, \frac{\Omega_{\rm b}h^2}{\sqrt{\Omega_{\rm m}h^2}} \n
&\times& \int_{z_1}^z dz (1+z)^{1/2} \, \Delta x(z)
\label{delta_tau}
\eea
with $H_{100}$ = 100 km/s/Mpc 
and $\Delta x = x_{\rm ion} - x_{\rm std}$, 
where $x_{\rm std}$ denotes the standard recombination 
history (i.e. without dark matter annihilation, 
and $m_{\rm s} = n_{\rm s}$). 
We have ignored dark energy, and therefore, the above equation 
holds true for $z_1 \gg 0$. 

We see that even small changes in $x_{\rm ion}$ at high redshifts 
can boost the total optical depth due to the $\sqrt{1+z}$ term. 
One may also hope to constrain dark matter annihilation by measuring the spectral distortion of the CMB, quantified by the Compton $y$ parameter \cite{chluba,pixie}:
\bea
y &=& \int d\tau \,\frac{k_{\rm b}\left[ T(z) - T_\gamma(z) \right ]}{m_{\rm e}c^2} \n
y-y_{\rm std} &\approx& \frac{c \sigma_{\rm T} \left( \rho_{\rm crit}/h^2 \right )}{H_{100} \, \bar m} \, \frac{\Omega_{\rm b}h^2}{\sqrt{\Omega_{\rm m}h^2}} \, \frac{k_{\rm b}}{m_{\rm e}c^2}  \n
&\times& \int_{z_1}^z dz (1+z)^{1/2} \, \left[ x_{\rm std} \Delta T + \Delta x (T_{\rm std}-T_\gamma) \right ], \;\;\;\;\;\;\;\;\;\;
\eea
where $\Delta T(z) = T(z)-T_{\rm std}(z)$, and as before, $T_{\rm std}$ represents the gas temperature in the standard $\Lambda$CDM scenario.

Fig. \ref{fig3}(a) shows the excess optical depth due to dark matter annihilation with $m_{\rm s}$ = 1.1, 1.3, and 1.5, with dark matter mass $m_\chi$ = 100 GeV, and concentration parameter $c_{200}$ = 5.  Assuming the total optical depth measured by Planck $\tau \approx 0.066$ \cite{newplanck_cosm} and full ionization up to $z=6$, we find that the excess optical depth $\Delta\tau \lesssim 0.026$, which excludes large values of $m_{\rm s}$. The Compton $y$ parameter is less constraining because it is weighted towards large $z$ when the gas temperature is close to the CMB temperature due to efficient  Compton scattering. The planned {\scriptsize PIXIE} mission can constrain $|y| < 2 \times 10^{-9}$ \cite{pixie}, and may exclude very large values of $m_{\rm s}$
at the relevant length-scales.

Fig. \ref{fig4} shows the evolution of the ionization fraction (Panel a)
and the gas temperature (Panel b), for different values of $m_{\rm s}$. 
We assume $f\sig/m_\chi = 1/100$ pb$\times c/$GeV for the figure. 
The black curve is plotted for the standard $\Lambda$CDM, i.e. ignoring 
dark matter annihilation. The red curve is plotted for the standard 
power law, but with accounting for dark matter annihilation. 
The green, blue, and magenta curves are the results 
for $m_{\rm s}$ = 1.1, 1.3, and 1.5, respectively. 
The effect on the CMB $TT$ and $EE$ power spectra is shown 
in Panels (c) and (d). 
For large $m_{\rm s}$, significant damping is caused on the $TT$ power
spectrum, as clearly seen in Panel (c).
Also plotted in (c) are the $TT$ power measurement from Planck. 
It might appear that large values of $m_{\rm s}$ are already excluded 
at high significance. However, the amplitude of the CMB power spectrum 
is determined by $\sim A_{\rm s} \exp-2\tau$. 
While the optical depth 
$\tau$ is increased by ionization by dark matter 
annihilation, the effect is almost fully degenerate with the amplitude 
of the primordial curvature power spectrum $A_{\rm s}$, 
except on very large scales that were outside the horizon 
at the time of particle annihilation. 
Therefore, the $TT$ power spectrum alone cannot be used to 
place constraints on dark matter annihilation, but
the degeneracy is broken by using information of the CMB polarization, 
because Thomson scattering causes a \emph{boost} in the large 
angle polarization power spectrum. A second technique to 
break the degeneracy is through the measurement of gravitational lensing 
of the CMB by large scale structure. The Planck experiment has recently 
measured the lensing potential at the 40$\sigma$ level \cite{newplanck_lens}. 
Measurement of the gravitational lensing of the CMB places constraints 
on the combination $\sigma_8 \Omega^{0.25}_{\rm m} = 0.591 \pm 0.021$. 
Since other constraints exist for $\Omega_{\rm m}$, the measurement of 
gravitational lensing places a bound on $\sigma_8$, and hence on $A_{\rm s}$. 
Gravitational lensing thus breaks the degeneracy between $A_{\rm s}$ and $\tau$. 
The recent results from Planck give us $\log (10^{10} A_{\rm s}) = 3.064\pm0.023$, 
and $\tau = 0.066 \pm 0.012$, which we can now be used to
place bounds on the root mean square mass fluctuation $\sigma_{\rm max}$, 
and hence on the power law index $m_{\rm s}$. 

We consider a number of models with different $m_{\rm s}$
and calculate $\sigma_{\rm max}$.
We also compute the corresponding ionization history and the CMB 
power spectra using a modified version 
of the {\scriptsize CAMB} software, assuming 
$p_{\rm ann} = f\sig/m_\chi = (1/100)$ pb$\times$c/GeV. 
For each model, we fit the theoretical power spectra to the observed $TT$ 
power spectrum, by modifying the quantity $A_{\rm s} e^{-2\tau}$.  
By means of Montecarlo simulations, we obtain a bound on the 
combination $A_{\rm s}e^{-2\tau} = 1.872 \pm 0.101$.  
The $2\sigma$ upper bound on $A_{\rm s}e^{-2\tau}$ results 
in a corresponding bound on the standard deviation of mass 
fluctuations: $\sigma_{\rm max} < 100$ at the 2$\sigma$ level. 
This finally translates to a bound on the power law index: 
$m_{\rm s} < 1.43 (1.63)$ for $k_{\rm p}$ = 100 (1000) $h$/Mpc.

\section{Conclusions}
We have proposed a new probe of the matter 
power spectrum on very small scales, 
through the effect of dark matter annihilation on the thermal 
evolution and on the ionization history of the inter-galactic gas. 
We have considered a simple modification to the standard power law, 
for the primordial power spectrum: 
$P_{\rm prim}(k) \sim k_p^{n_{\rm s}} (k/k_p)^{m_{\rm s}}$, 
where $m_{\rm s} \geq n_{\rm s}$. Such a power law is acceptable 
provided $k_{\rm p}$ is large enough, e.g. $k_{\rm p} > 10$ $h$/Mpc.
The form of the root mean square mass fluctuation $\sigma$ 
has been calculated as a 
function of the power law index $m_{\rm s}$. 
The maximum value 
$\sigma_{\rm max} = \sigma(M_{\rm min})$ varies significantly with $m_{\rm s}$. 
One may expect a $1-\sigma$ fluctuation to enter the non-linear regime 
when $1+z \approx \sigma_{\rm max} / \delta_{\rm c} D(0)$. 
We have then computed the filling fraction (the fraction of dark matter
bound in nonlinear halos) as a function of redshift.
For large $m_{\rm s}$, there are many orders of magnitude 
more halos at $z>100$. 
For the standard power spectrum power law, halos are only 
important for $z \lesssim 50$. On the other hand, 
when $m_{\rm s} \sim 1.5$, halos provide the dominant contribution
to total dark matter annihilation rate even at $z=300$.

We derived an explicit expression for the energy injected 
per unit gas atom per unit time at a redshift $z$. 
A large contribution from dark matter halos at $z \sim 300$ can 
significantly alter the spectrum of CMB anisotropies. 
We computed the CMB power spectra using the 
{\scriptsize CAMB} code, for a fiducial dark matter
annihilation cross-section of 
$p_{\rm ann} = f\sig/m_\chi = (1/100)$ pb$\times$c/GeV. 
We used the Planck (2015) $TT$ power 
spectrum data to test theoretical models.
The current data already excludes a root mean square fluctuation 
$\sigma_{\rm max} = \sigma(M_{\rm min}) \gtrsim 100$ at the 2$\sigma$ level. 
The bound on $\sigma_{\rm max}$ may be expressed as a constraint 
on the power law index on small scales: 
We exclude $m_{\rm s} > 1.43 (1.63)$ for $k_{\rm p}$ = 100 (1000) $h$/Mpc.

\acknowledgments{A.N. acknowledges funding from the Japan Society for the Promotion of Science (JSPS) and the Kavli Institute for the Physics and Mathematics of the Universe (IPMU). A.N. is grateful to Queen's University and the University of Pennsylvania for hospitality and funding. The authors thank David Spergel for fruitful discussions, and also for suggesting the use of the Planck measurements of the optical depth and the fluctuation amplitude. NZ is grateful for the hospitality of David Spergel
and the Department of Astrophysical Sciences at Princeton University. NZ's visit was supported by the University of Tokyo-Princeton strategic partnership
grant.}

\bibliographystyle{revtex}
\bibliography{references}

\end{document}